\documentclass[runningheads,a4paper]{llncs}

\usepackage{amssymb}
\usepackage{graphicx}
\usepackage{caption, booktabs}
\usepackage{subcaption}
\usepackage{listings}
\usepackage{xcolor}
\usepackage{algorithm2e}
\usepackage{hyperref}

\usepackage{url}
\urldef{\mailsa}\path|{alfred.hofmann, ursula.barth, ingrid.haas, frank.holzwarth,|
\urldef{\mailsb}\path|anna.kramer, leonie.kunz, christine.reiss, nicole.sator,|
\urldef{\mailsc}\path|erika.siebert-cole, peter.strasser, lncs}@springer.com|    
\newcommand{\keywords}[1]{\par\addvspace\baselineskip
\noindent\keywordname\enspace\ignorespaces#1}

\begin{document}


\definecolor{forestgreen}{RGB}{34,139,34}
\definecolor{orangered}{RGB}{239,134,64}
\definecolor{darkblue}{rgb}{0.0,0.0,0.6}
\definecolor{gray}{rgb}{0.4,0.4,0.4}
\definecolor{codegreen}{rgb}{0,0.6,0}
\definecolor{codegray}{rgb}{0.5,0.5,0.5}
\definecolor{codepurple}{rgb}{0.58,0,0.82}
\definecolor{backcolour}{rgb}{0.95,0.95,0.92}

\lstdefinestyle{XML} {
    language=XML,
    extendedchars=true, 
    breaklines=true,
    breakatwhitespace=true,
    emph={},
    emphstyle=\color{red},
    basicstyle=\ttfamily,
    columns=fullflexible,
    commentstyle=\color{gray}\upshape,
    morestring=[b]",
    morecomment=[s]{<?}{?>},
    morecomment=[s][\color{forestgreen}]{<!--}{-->},
    keywordstyle=\color{orangered},
    stringstyle=\ttfamily\color{black}\normalfont,
    tagstyle=\color{darkblue}\bf,
    morekeywords={attribute,xmlns,version,type,release},
    otherkeywords={attribute=, xmlns=},
}

\lstdefinelanguage{PDDL}
{
  sensitive=false,    
  morecomment=[l]{;}, 
  alsoletter={:,-},   
  morekeywords={
    define,domain,problem,not,and,or,when,forall,exists,either,
    :domain,:requirements,:types,:objects,:constants,
    :predicates,:action,:parameters,:precondition,:effect,
    :fluents,:primary-effect,:side-effect,:init,:goal,
    :strips,:adl,:equality,:typing,:conditional-effects,
    :negative-preconditions,:disjunctive-preconditions,
    :existential-preconditions,:universal-preconditions,:quantified-preconditions,
    :functions,assign,increase,decrease,scale-up,scale-down,
    :metric,minimize,maximize,
    :durative-actions,:duration-inequalities,:continuous-effects,
    :durative-action,:duration,:condition
  }
}

\lstdefinelanguage{Srv}{
keywords = [1]{bool, uint8, int32, uint64, float32, float64, string, Header, Point, Quaternion, time},
comment=[l]{\#}
}

\lstdefinestyle{mystyle}{
  backgroundcolor=\color{backcolour},   commentstyle=\color{codegray},
  keywordstyle=\color{codegreen},
  numberstyle=\tiny\color{codegray},
  stringstyle=\color{codepurple},
  basicstyle=\ttfamily\footnotesize,
  breakatwhitespace=false,         
  breaklines=true,                 
  captionpos=b,                    
  keepspaces=true,                 
  numbers=left,                    
  numbersep=5pt,                  
  showspaces=false,                
  showstringspaces=false,
  showtabs=false,                  
  tabsize=2
}
\lstset{style=mystyle}

\mainmatter  

\title{Virtual Reality in University Teaching: Experiences from a Computer Science Seminar}

\titlerunning{Virtual Reality in University Teaching}

%
%
%

\author{Enes Yigitbas}

\authorrunning{Enes Yigitbas}

\institute{Paderborn University\\ Zukunftsmeile 2, 33102 Paderborn, Germany\\
\email{enes@mail.upb.de} 
}

%
%

\toctitle{Lecture Notes in Computer Science}
\tocauthor{Authors' Instructions}
\maketitle

\begin{abstract}
Due to the corona pandemic, numerous courses were held using digital solutions in order to be able to continue teaching. Conventional collaboration tools (Zoom, Big Blue Button, etc.) were used in particular to digitally map a synchronous session for teaching and learning purposes. While these conventional collaboration tools offer a solid basis for communication between learners and teachers, aspects such as presence or a realistic type of interaction are neglected. In this work, we report on the experiences from a computer science seminar where virtual reality (VR) technology was used as an alternative solution for teaching and group work. The benefits of VR compared to conventional collaboration tools were examined using questionnaires and interviews with the participants. On the one hand, the results show the high potential of VR to increase the clarity and experienceability of learning content and to promote cooperation through social presence. On the other hand, the use of VR brings with it some technical and organizational difficulties that should be taken into account in the didactic implementation.
\end{abstract}

\section{Introduction}\label{sec:intro}

Digital devices and media are increasingly being used for learning and educational purposes these days. This was already evident in the period 1997-2006 when networked computers were used extensively for collaborative learning, and in the period 2007-2016 when so-called online digital learning became widespread \cite{radianti2020systematic}. In recent years, virtual reality (VR) technology has been actively integrated into education, teaching, and training in various application domains \cite{DBLP:books/sp/22/DBGJ2022}. Although VR is not new and was defined as "a real or simulated environment in which a perceiver experiences telepresence" \cite{steuer1993defining}, recent developments in display technology and computer graphics have made VR more affordable and available to a wider spectrum made accessible by humans. The latest VR head-mounted displays (HMDs) such as Valve Index or Oculus Quest 2 allow users to immerse themselves in a realistic virtual world. The potential of VR technology for education is compelling according to experts and according to some experts like Sol Rogers VR is the learning aid of the 21st century \cite{rogers2019virtual}. This is mainly due to the fact that learning activities and practical experiments, which are often too expensive, too dangerous, or simply too time-consuming in the real-physical learning space, can be replaced by virtual experiences. Based on Pirker et al. \cite{pirker2020virtual}, translating these experiences into immersive virtual realities can be of great benefit to learners, as the sense of being physically, socially, and personally present in the virtual environment contributes to the learning process.

In view of these promises of VR technology, the question arises as to what the advantages and disadvantages of using VR are compared to classic collaboration tools such as Zoom or Big Blue Button. Due to the corona pandemic and the associated restrictions, numerous courses on digital solutions were held in order to overcome the challenges of classroom teaching and to offer alternative ways of being able to continue teaching. In particular, the classic collaboration tools mentioned were increasingly used to digitally map a synchronous session for teaching and learning purposes. While these offer a solid basis for communication and interaction between learners and teachers, aspects such as presence (i.e. the feeling of "being there") or a realistic type of interaction (e.g. through natural movement, gestures, etc.) are neglected. In order to explore an alternative solution to this problem, a new teaching concept was developed in a computer science master's seminar "Mixed Reality Software and Technology (XRST)" (SS 2021) at the University of Paderborn, in which VR technology is used as the primary medium for the course and the associated group work. In the course of the seminar, various plenary sessions (workshops, presentations, discussions, etc.) and group work were carried out using a VR learning room. Using questionnaires and interviews with the participants, the user-friendliness, interaction, collaboration and learning support of VR were examined in comparison to conventional collaboration tools. In this work, the advantages and disadvantages of using VR in university teaching are analyzed and discussed using the example of a computer science seminar. The focus is on promoting collaboration, interaction, and learning support from VR compared to conventional collaboration tools. In this context, evaluation results based on the questionnaires and interviews are presented and critically reflected upon.

This work is structured as follows: Chapter 2 describes the basic concepts on the subject of Virtual Reality. Chapter 3 deals with related work in the context of VR in university teaching. Chapter 4 then presents the research questions that are to be answered in the teaching research project. Chapter 5 explains the new teaching concept and the course of the event, which was primarily based on VR. The evaluation of the event with reference to the research questions is discussed in Chapter 6. A discussion and critical reflection of the evaluation results and findings can be found in Chapter 7. The article ends with a summary and outlook in Chapter 8.

\section{Background and Related Work}\label{sec:background}

This chapter described relevant background information. First, a brief introduction to Virtual Reality technology is given. Then, VR-based collaboration tools are presented, with a particular focus on the VR collaboration platform "Spatial" used in the seminar under consideration.

\subsection{Introduction to Virtual Reality Technology}

Virtual Reality (VR) is the representation and simultaneous perception of reality and its physical properties in a real-time, computer-generated, interactive virtual environment. According to Biocca and Delaney \cite{biocca1995immersive}, VR can be defined as "the total sum of hardware and software systems attempting to perfect an all-encompassing, sensory illusion of presence in another environment." Immersion, presence, and interactivity are considered core features of VR technology (\cite{ryan2015narrative}, \cite{walsh2002virtual}). According to Slater and Wilbur \cite{slater1997framework}, immersion is defined as "a perception of being physically present in a non-physical world by surrounding users of a VR environment with images, sound or stimuli" so that they feel to actually be "there". Immersion thus describes the embedding of the user in the virtual world. The perception of oneself in the real world is reduced and the user feels more like a person in the virtual world. The more immersive a VR experience is, the more realistic it feels to the user. Presence is defined as "the subjective experience of being in one place or environment, even when physically located elsewhere" \cite{witmer1998measuring}. Finally, the term interactivity can be described as the extent to which users can change the VR environment in real-time \cite{steuer1993defining}. VR technology is becoming increasingly sophisticated and cost-effective and can be applied to many areas such
as training~\cite{DBLP:conf/vrst/YigitbasJSE20}, prototyping (e.g., \cite{DBLP:conf/hcse/JovanovikjY0E20}, \cite{DBLP:conf/vl/YigitbasKGE21}), robotics \cite{DBLP:conf/seams/YigitbasKJE21}, education (e.g., \cite{DBLP:conf/mc/YigitbasTE20}, \cite{DBLP:conf/models/YigitbasSBGE22}), healthcare (e.g., \cite{DBLP:conf/mc/YigitbasHE19}, \cite{yigitbas2022technical}), finance~\cite{yigitbas2022development}, or even software modeling (e.g., \cite{DBLP:conf/models/YigitbasGWE21}, \cite{yigitbas2022design}).

While VR realizes the idea of a virtual world in which the user feels present through the manipulation of their senses and is completely immersed in a virtual world without having any relation to the outside world, augmented reality (AR) is an extension of reality. With augmented reality, the real-physical world is enriched with digital information such as text, 3D objects, images, etc. \cite{azuma1997survey}. Similar to VR, AR has already been applied in various application domains such as robot programming (e.g., \cite{DBLP:conf/interact/YigitbasJE21}, \cite{DBLP:conf/eics/KringsYBE22}), product configuration (e.g., \cite{DBLP:conf/hcse/GottschalkYSE20}, \cite{DBLP:conf/hcse/GottschalkYSE20a}), planning and measurements \cite{DBLP:conf/eics/EnesScaffolding} or for realizing smart interfaces (e.g., \cite{DBLP:conf/eics/KringsYJ0E20}, \cite{DBLP:conf/interact/YigitbasJ0E19}). While AR also has great potential for teaching and learning purposes, VR technology is the main focus of this work.

As the already introduced definition of VR suggests, special hardware and software are required for an immersive VR experience. In order to create a feeling of immersion, special input and output devices called VR headsets are required to display virtual worlds. The Oculus Quest 2 was used to carry out this work. This mobile, wireless standalone VR headset consists of a head-mounted display (HMD) and two hand controllers. When using the HMD, two images are generated and displayed from different perspectives in order to convey a spatial impression. The HMD is also used to track head movements. This tracking serves to map the movements, for example by changing the viewing angle, in VR. Analogously, inputs can be made via the hand controller and hand and arm movements can be tracked, which overall enables natural control of VR applications. In addition to special hardware, special software is also required to support a VR experience. To create virtual reality, you need software specially developed for this purpose. These programs must be able to calculate complex three-dimensional worlds in real-time, in stereo (separately for the left and right eye). Special software development tools (SDKs) and game engines are available for this purpose to support the development of complex virtual 3D worlds. In practice, however, there are already existing VR applications and tools for many areas of application that simplify work in the respective specialist domain.

\subsection{VR Collaboration Tools}

A collaboration tool is a digital tool for working together in a team. This involves communication, joint project work, and topics such as the collaborative processing of documents. Similar to the traditional collaboration tools (Big Blue Button (BBB), Teams, Zoom, etc.), numerous VR collaboration tools enable a synchronous session in Virtual Reality. Engage VR\footnote{https://engagevr.io/}, Altspace VR\footnote{https://altvr.com/} and Spatial\footnote{https://spatial.io/} are some of these VR collaboration tools among many others. Based on a technological evaluation, the VR collaboration platform "Spatial" was selected to carry out this work, which is described in more detail below.

"Spatial" was originally developed for companies. Customers include well-known corporations such as Mattel, Nestlé, and Pfizer. In the wake of the Corona crisis, the start-up opened its platform to the masses by making its services, including enterprise functions, available to end users without restrictions and free of charge.

In order to support collaboration in VR as realistically as possible, "Spatial" provides a number of features. Based on a selfie photo or profile picture, users can have a realistic 3D avatar created of themselves, which characterizes the respective user as a digital twin during a virtual session. This avatar is brought to life during a VR session by reflecting the user's movements onto the avatar. In addition, different prefabricated virtual rooms can be selected for a virtual session (office, seminar room, lecture hall, campfire, etc.). Furthermore, digital elements can be created in the selected virtual space by adding 3D objects, photos, images from external sources (hard disk or web). Other features include free-hand drawing in VR, searching in the VR web browser or the integration of other services such as Slack or Google Drive. In addition, the work artifacts (text, images, 3D objects, etc.) can be viewed and edited together with other participants. In addition to the avatars, communication options via "spatial audio" are available for collaboration, or the option of also seeing each other via the webcam. The idea of "Spatial Audio" is to enable auditory communication as in presence, i.e. the closer you are to a person virtually, the better you can hear the person. "Spatial" aims at a collaboration in VR over different device classes. In addition to typical VR headsets such as Oculus Quest or Valve Index, AR glasses such as Hololens, Magicleap or Nreal are also supported to access the virtual collaboration environment. In addition, a web interface is provided for non-immersive device classes such as desktop PCs, laptops or smartphones, which can be used to show the virtual collaboration environment as a 3D space on the screen. This form of access or use of "Spatial" is referred to in this article as desktop VR (DVR). While the use of "Spatial" via an HMD enables an immersive experience where the collaboration space and other participants can be experienced as realistically as possible, the use via non-immersive device classes (DVR) is a 3D representation that is based on a 2D screen.

\section{Initial Situation: Use and Evaluation of VR Technology in University Teaching}

The early adoption of VR-based solutions for teaching and educational purposes began with desktop-based 3D virtual environments. For example, a very popular virtual world called “Second Life” \cite{esteves2011improving} has been introduced to create digital twins of real-world locations in which users, represented digitally in the form of avatars, actively participate in realistic activities that stimulate learning. Although these desktop-based virtual environments (DVR) cannot provide a fully immersive experience, their photorealistic computer graphics have been shown to improve learner engagement \cite{dickey2003teaching}.

A survey of VR usage in teaching and education \cite{radianti2020systematic} shows that VR is most commonly used in the following educational application areas: Engineering (24\%), Computer Science (10\%), Astronomy (7\%), and Biology (5\%). Most of the existing approaches to using VR in teaching are very specific or limited to single-user scenarios. In this context, \cite{hagge2021student} presents, for example, experiences with VR in a geography lecture in which only one student uses a VR headset and the rest of the class watches the streamed VR video.

Similarly, the authors in \cite{garcia2019nomad} have developed a VR environment in which individual participants can engage in a virtual chemistry lab. In addition, projects such as the Erasmus+ project "Virtual Reality in Higher Education: Application Scenarios and Recommendation"\footnote{https://www.uni-due.de/proco/erasmusplus.php} (2018-2020) or the launched project "AR/VR.NRW"\footnote{https://www.hshl.de/forschung-unternehmen/forschungsprojekte/forschungsprojekte-im-themenfeld-materialwissenschaften/ar-vr-nrw/} underline the importance and potential of VR for university scenarios.

However, existing approaches in this area do not adequately cover the aspects of multi-user collaboration and interaction in a VR setting and their impact on the learning outcome. In this context, it can be observed that VR has so far mostly been part of the experimental and development work and has not been used regularly in teaching. In particular, pilot projects in which VR is used throughout a semester and student perception is analyzed in order to evaluate the user-friendliness of this medium in comparison to traditional collaboration tools such as Big Blue Button or Zoom are missing in teaching at universities.

\section{Research Questions}

With the advent of VR technology, it is possible to involve teachers and learners in a virtual classroom where immersion and presence, key factors of VR, can contribute to learning processes by creating a sense of "being there". In addition to immersion and presence, VR technology has great potential to create collaborative learning experiences for modern educational institutions \cite{zheng2018affordances}. Although the advancement of VR technology expands collaborative learning, related research (see Section 3) on VR-based collaborative learning is still limited. Therefore, it is important to investigate how the use of VR affects collaborative learning and how VR influences collaboration and interaction in learning processes. From this we derive the following research questions and hypotheses for this work:

(RQ1) How do the students perceive the use of VR in terms of user-friendliness compared to conventional collaboration tools? How does the use of VR technology differ in terms of usability when accessed via HMDs or non-immersive device classes (desktop, laptop, tablet, etc.)? My hypothesis in terms of usability is that traditional collaboration tools have higher acceptance and generally perform better because many users are familiar with them. With regard to the distinction between VR via HMDs and non-immersive device classes, I expect a significantly higher level of user-friendliness for the immersive version where HMDs are used.

(RQ2) How do the students perceive the use of VR in terms of collaboration compared to conventional collaboration tools? My hypothesis is that the use of VR enables a more natural (realistic) form of collaboration and that this is positively reflected in aspects such as motivation and fun. On the other hand, it could be that the efficiency and effectiveness of collaboration suffer since VR is a new medium for many, where the input and output options are not familiar when used.

(RQ3) How do the students perceive the use of VR in terms of interaction compared to conventional collaboration tools? Here is my hypothesis that VR increases the perception of presence, since the depiction of the participants as 3D avatars and the enabling of realistic forms of interaction support an immersive experience that is not possible with conventional collaboration tools.

(RQ4) How do students perceive the use of VR in terms of learning support compared to conventional collaboration tools? Here it is to be expected that the use of VR could be particularly useful for learning processes where visual and spatial perception is required and should be underpinned by interactive learning steps.

\section{Teaching Concept and Course of the Event}

The basis of the teaching concept was the original seminar concept of the Institute for Computer Science. According to the description of the module manual, the participants work on a topic in a seminar, which is presented in a lecture followed by a discussion and a written elaboration. The focus here is on the independent or joint development of research-related sub-areas of computer science. In the master's computer science seminar "Recent Advances in Mixed Reality Software and Technology", which deals with the topics of augmented and virtual reality and their basics and fields of application, virtual reality was used for the first time in the 2021 summer semester as the primary medium for conducting the event carried out. At the university, there is (to the best of our knowledge and belief) no comparable format in which the advantages and disadvantages of using VR in terms of collaboration, interaction, and learning support have been analyzed in detail.

\begin{figure}
    \includegraphics[width=1\linewidth]{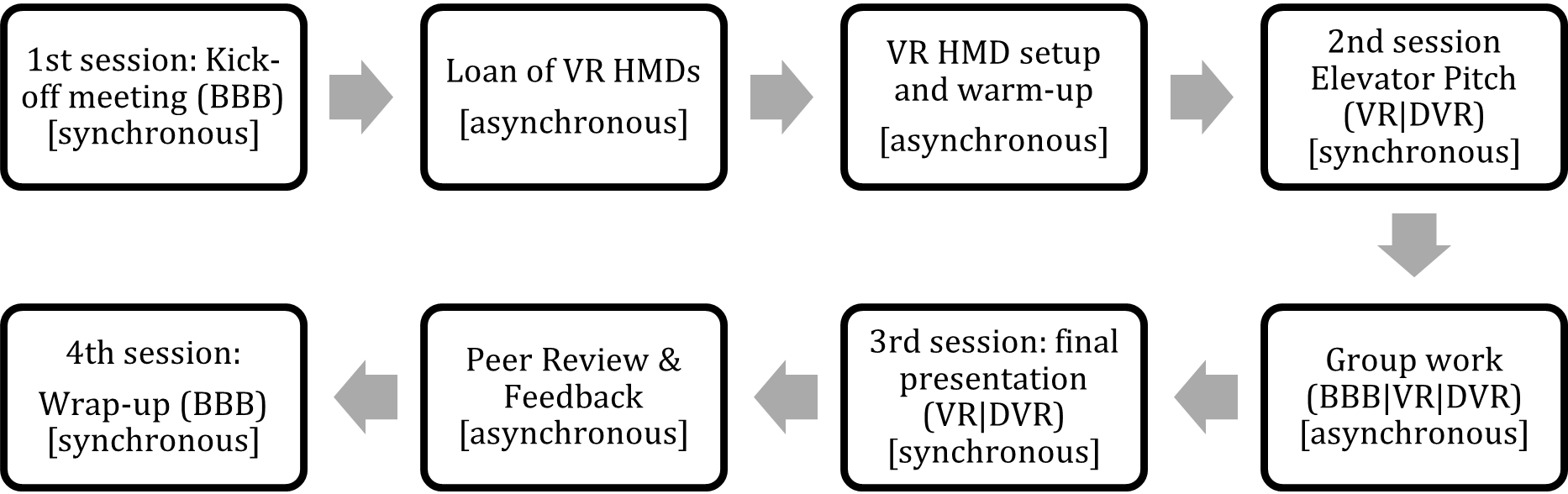}
    \caption{Seminar Schedule}
    \label{fig1}
\end{figure}

Twelve participants were registered for the course. The seminar was designed and carried out as follows. At the content level, the participants had the task of selecting a topic from a list of offered topics in the context of augmented and virtual reality and creating an elaboration and a presentation on this topic. The seminar was organized as shown in Figure \ref{fig1}.

First, the seminar started with a kick-off meeting via BBB. Since this was the very first meeting and most of the participants did not have VR headsets, it was conducted via BBB. The main purpose of this meeting was to encourage the participants to get to know each other and to provide them with organizational information and a list of seminar topics. In addition, the participants were informed about the SoTL project \cite{huber2014scholarship}.

\begin{figure}
    \includegraphics[width=1\linewidth]{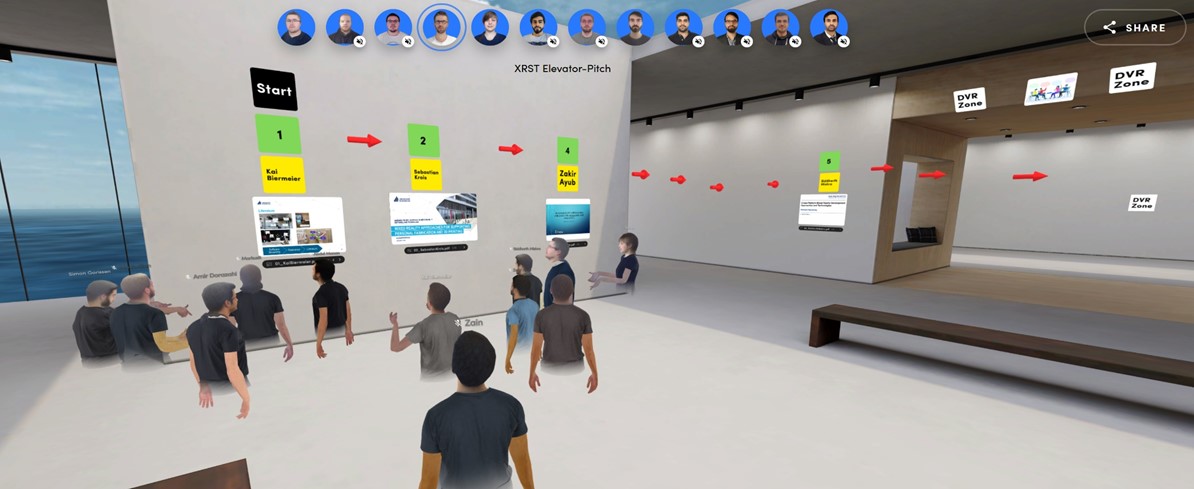}
    \caption{Screenshot from the Elevator Pitch in VR}
    \label{fig2}
\end{figure}

After the kick-off meeting, the VR participants in the seminar were randomly selected as the number of available VR headsets (Oculus Quest 2) was limited to six.
After borrowing the VR headsets, there was a VR setup and warm-up phase, which served to set up the VR environment and test it with the participants so that everything worked technically flawlessly.
The 2nd session was then carried out in VR, with half of the participants connecting to the VR learning room "Spatial" via the VR headsets (hereinafter referred to as VR users) and the other half connecting via their desktops or laptops based on the web application of "Spatial" (hereinafter referred to as DVR users). In this session, there was an elevator pitch, where the participants were supposed to give a short (max. 3 min.) presentation on their seminar topic. A screenshot from the "Spatial" VR environment during the elevator pitch is shown in Figure \ref{fig2}.

The participants should then form groups for the upcoming work phase in the seminar. The method of speed dating in VR was used to form the groups, where different zones were provided. When the groups were formed, the rule was that three different groups should be created where BBB users, VR users, and DVR users could meet exclusively. The aim of this rule was that the participants with different media uses did not mix with each other, thereby enabling a better comparison between the BBB, VR, and DVR users in the seminar.

After the groups had been formed in the 2nd session (3 VR groups of 2 participants each, a DVR group of 2 participants, and a BBB group of 3 participants) the group work phase began, in which each group shared a presentation of the respective seminar topics.

For the creation of the final presentation, the task of the VR and DVR participants was to create a VR environment to present their results, while the BBB teams could use all media formats allowed in BBB. When carrying out the group work, there was a strict requirement that each group should use the respective medium BBB, VR or DVR for the group exchange as a meeting. After the end of the group work phase, there was the next session in VR, where the final presentations of the groups were presented.

\begin{figure}
    \includegraphics[width=1\linewidth]{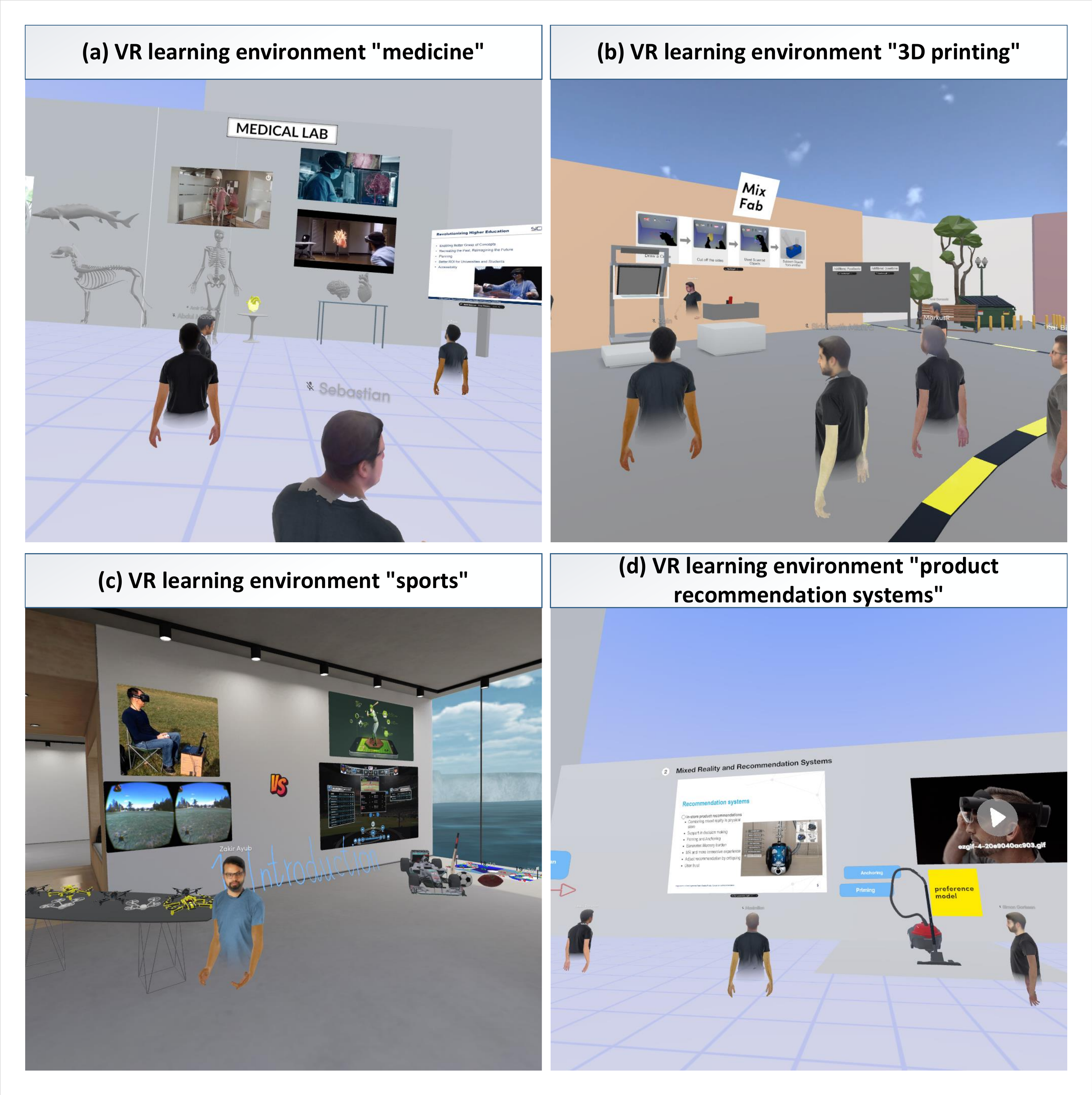}
    \caption{Screenshots from the VR environments during the final presentation}
    \label{fig3}
\end{figure}

Very creative and interesting VR environments were created (see Figure \ref{fig3}) to present the results of the individual and group work. Some screenshots from the VR environments are shown in Figure \ref{fig3} as an example. In Figure \ref{fig3} (a), for example, a group developed a VR learning environment for the medical field, where body parts of humans and animals can be explored and learned in a virtual world. In Figure \ref{fig3} (b), another group explained how 3D printing and additive manufacturing works in VR. Another group presented in VR how augmented reality-based drone control can be used in the context of sports (see Figure \ref{fig3} (c)). Finally, another group created a VR learning environment, where the topic of product recommendation systems based on 3D product models is explained in VR (see Figure \ref{fig3} (d)).

After completing the final presentations in VR, there was a peer review and feedback phase where the participants were asked to read and review a seminar draft by another participant. For this purpose, the well-known conference management system "Easychair" was used to familiarize the participants with the organization and implementation of conferences and review procedures. Finally, the seminar ended with the 4th session "Wrap-up". This last session took place via BBB in order to gather all participants again in the same medium and to offer them the opportunity to give final feedback on the content and organization of the seminar.

\section{Evaluation}

In this chapter, the results of the use of virtual reality in the underlying seminar are examined in more detail. For this purpose, the data collected during the seminar is considered. Within the analysis, the research questions defined in Chapter 4 are answered.

\subsection{Data Collection}

In the course of the event, various data were collected in order to analyze the respective research questions. First, in the “1st Session: Kick-off meeting (BBB)” (see Figure 1) a questionnaire was filled out by the twelve participants, which dealt with the usability, quality of collaboration and interaction, as well as the participants' perception of learning via BBB. There was a similar questionnaire after the “2nd Elevator Pitch Meeting” (see Figure 1) for the VR and DVR media. There were six VR and five DVR participants. As a supplement, there were additional questions about the perception of presence in order to be able to analyze the differences between VR and DVR with regard to immersion. The questionnaires mentioned also contained open questions on the general advantages and disadvantages of the respective media (BBB, VR, DVR). In addition, the participants were asked to write minutes of their group work meetings in the respective media, which were also taken into account in the analysis. This form of approach to teaching research was chosen to provide a comprehensive view of the use of VR by combining a quantitative and qualitative analysis. The quantitative approach based on the evaluation of the questionnaires enables a comparative analysis of the media examined with regard to the relevant criteria. The open questions and minutes of the group work opened up the possibility of assessing the user-friendliness and the strengths and weaknesses of the respective media using a qualitative form of analysis.

\subsection{Usability and Presence Perception}

\begin{figure}
\centering
    \includegraphics[width=0.8\linewidth]{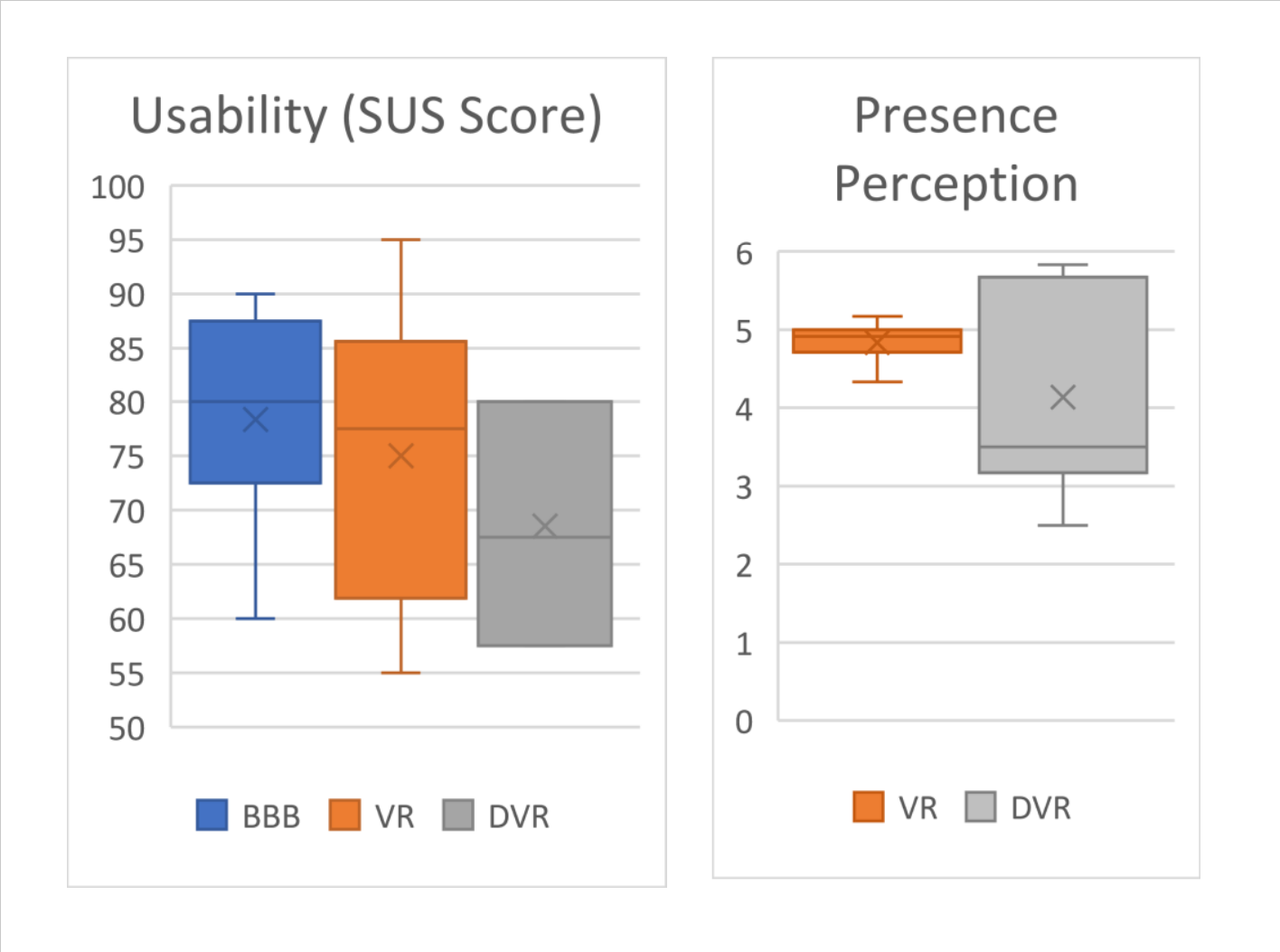}
    \caption{Usability and presence perception of the different media used}
    \label{fig4}
\end{figure}

The System Usability Scale (SUS) questionnaire \cite{lewis1995ibm} was used to analyze the usability of the media used in the seminar (BBB, VR, DVR). The SUS questionnaire is a standardized questionnaire for assessing the user-friendliness of an interactive system. It consists of a total of 10 questions that are rated on a Likert scale from 1-5. Based on these answers, a SUS score between 0-100 can be determined, which is an indication of the usability of the interactive system as a whole. As can be seen in the box-plot diagram in Figure \ref{fig4} on the left, the average ease of use is highest for Big Blue Button (BBB) with a SUS score of around 78. The medium of virtual reality based on the "spatial" application (VR SUS score 75) takes second place in terms of user-friendliness, while the use of the same application via desktop or laptop (DVR SUS score: 68.5 ) performs worst. Based on the evaluation scheme according to Bangor et al. \cite{bangor2009determining}, the SUS scores for BBB and VR can be classified as "good" with minimal differences, while the SUS score of DVR can be classified as "ok". (A SUS score above 85 is expected for an “excellent” rating.) These observations are not surprising as most participants are very familiar with BBB and have been using it for a long time. On the other hand, according to their own statements, more than half of the VR participants were not familiar with the use of a VR HMD or the VR collaboration platform "Spatial". For this reason, the user-friendliness of VR should not be viewed as negatively, even if there is still room for improvement in some areas. However, the user-friendliness from the perspective of the DVR user is significantly worse, since the use of a VR application without a VR HMD does not result in great added value and the display in 3D alone is not sufficient to enable a realistic meeting. This is also underlined again based on the data collected with regard to the presence questionnaire. Here, too, an established questionnaire for the perception of presence was used by Witmer and Singer \cite{witmer1998measuring}. The results show that the average presence perception with VR is significantly better compared to DVR (see Figure \ref{fig4} right).

In summary, with a view to the research question (Q1), it can be said that the user-friendliness of established conventional tools such as BBB is very high and that most users can handle them well, which confirms the hypothesis. However, the use of virtual reality is associated with initial hurdles and difficulties, since most users need an introduction or training in the required hardware or software, which they are usually not completely familiar with at the beginning. Nevertheless, after overcoming this entry problem, VR can serve as a medium for a seminar to hold presentations, meetings, etc. It is important to emphasize that the availability of VR hardware is extremely important and access to VR via non-immersive devices such as desktop PCs, laptops or tablets does not offer a great perception of presence and therefore the potential of VR cannot fully unfold.

\subsection{Collaboration}

\begin{figure}
\centering
    \includegraphics[width=0.9\linewidth]{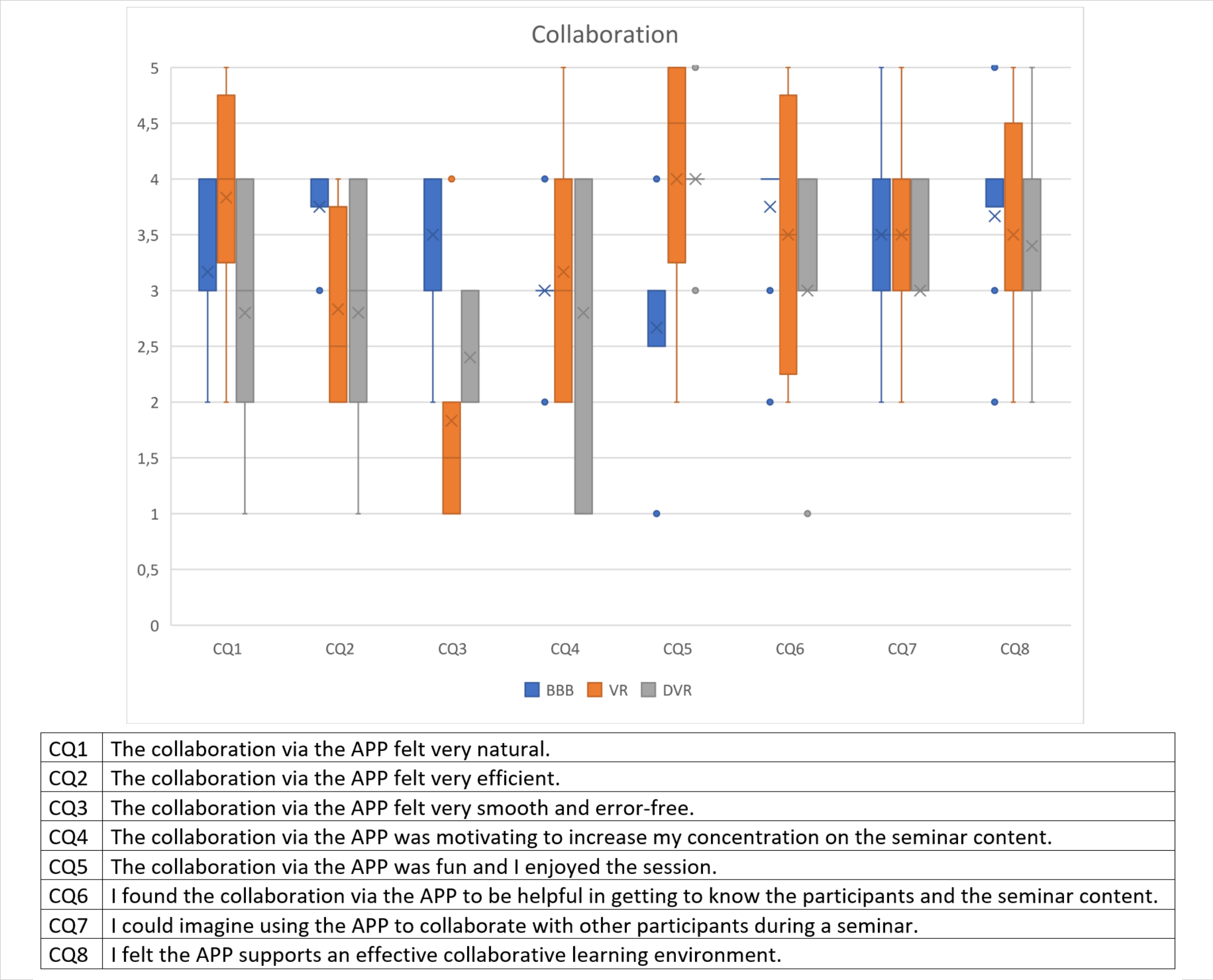}
    \caption{Evaluation results concerning collaboration}
    \label{fig5}
\end{figure}

To analyze the strengths and weaknesses of the respective media in terms of collaboration opportunities, a separate questionnaire was created based on literature research on the topics of "usability evaluation" and "collaboration in VR". This contains a total of eight questions or statements CQ1-CQ8, which are shown in Figure \ref{fig5} and were answered on a Likert scale from 1-5, similar to the SUS questionnaire. With regard to the first statement "CQ1 - The collaboration via the APP felt very natural" it quickly becomes apparent that VR achieves the best result on average. BBB is second in terms of natural collaboration, and DVR has the worst results on average. This can possibly be explained by the inherent properties of the modern VR interaction technology, which not only reproduces or simulates the virtual space realistically, but also makes it possible to experience and touch it naturally through movement, gesture control, etc. BBB and DVR perform relatively poorly here since they are based on a 2D-based user interface that is operated via mouse/keyboard.

However, if we look at collaboration in terms of the perceived efficiency (CQ2), it quickly becomes clear that the picture is changing. BBB achieves significantly better results here compared to VR and DVR, which perform similarly poorly. This can be explained by the fact that direct manipulation of the display and control elements is possible in classic 2D-based user interfaces such as BBB, while real control operations such as movements take longer in VR and DVR.

A similar picture can be seen with regard to the perceived effectiveness (CQ3) of collaboration, where BBB achieves the best and VR the worst rating on average. This result can be explained by the fact that precise input options in VR and DVR are only possible to a limited extent. In VR, inputs are usually made using the hand controllers, which are not yet as precise to use as a regular keyboard. In DVR you have the keyboard of the non-immersive device used, but the processing of 3D elements is error-prone and time-consuming.

Regarding the statement "CQ4 - The collaboration via the APP was motivating to increase my concentration on the seminar content" it becomes apparent that VR performs slightly better than BBB and DVR. This may be due to the increased perceptual presence described in Section 6.2.

With regard to the fun factor (CQ5) when conducting the seminar, it becomes clear that VR and DVR achieved significantly higher results on average than BBB. This may be explained by the fact that most were new to VR as a new technology while already familiar with BBB. The immersive and dynamic exploration and experience of the VR environment could also have contributed to the fun factor in VR being felt so positively.

Based on the CQ6 statement, it becomes clear that BBB was classified on average as a more helpful medium for getting to know the participants and seminar content. This could be related to the chat function, where participants can write to and contact each other directly. In VR/DVR, this was not directly possible in the "Spatial" collaboration platform used and one would have had to withdraw with one of the participants in order to be able to have a private conversation.

As statement CQ7 shows, most participants can imagine using BBB and VR as a medium for group work in the seminar. On the other hand, the rating from DVR was rather negative in this respect, which shows that many do not see this solution as the right alternative for collaboration in the seminar.
Finally, the evaluation of the CQ8 statement makes it clear that on average BBB is seen as the most effective collaborative learning environment among the three media compared, because on the one hand it is probably known and on the other hand it provides the necessary features for collaboration in a pragmatic and simple manner.

In summary, with a view to the research question (RQ2), it can be deduced that VR and DVR bring no real perceived improvement in comparison to conventional collaboration tools (BBB, Zoom, etc.) in terms of efficiency and effectiveness. However, it can be seen that VR can have added value if it is used in a targeted manner to increase the motivation of the participants and to promote playful learning through doing with a fun factor. Overall, with reference to the research question (RQ2), the hypothesis can also be confirmed here.

\subsection{Interaction}

\begin{figure}
\centering
    \includegraphics[width=0.9\linewidth]{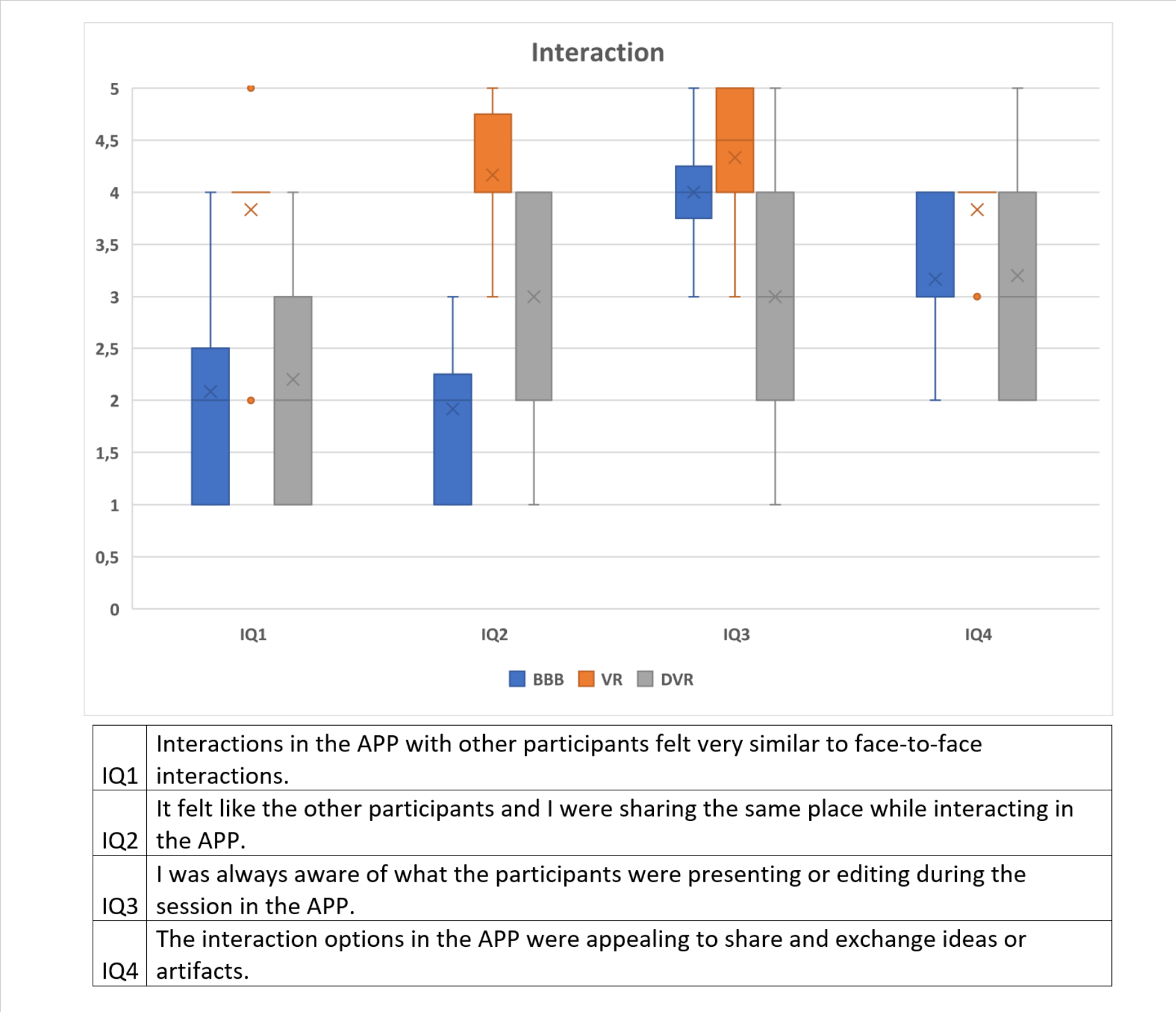}
    \caption{Evaluation results concerning interaction}
    \label{fig6}
\end{figure}

Analogously to the aspect of collaboration, a separate questionnaire was created to examine the interaction properties. This contains a total of four questions or statements IQ1-IQ4, which are shown in Figure \ref{fig6} and were answered on a Likert scale from 1-5, similar to the SUS questionnaire.

Regarding the statement "IQ1 - The interactions with other participants of the APP felt very similar to the personal (face-to-face) interactions.", on average, VR received a significantly higher rating than BBB and DVR. This can be explained, for example, with the 3D avatars that each person can create for themselves. By controlling this avatar, for example, with hand, arm and head movements, the impression of face-to-face interaction in VR is conveyed more. This is only possible to a limited extent in DVR since the avatar can only be moved forwards, backwards, left, and right, but other forms of movement are not possible. In BBB this is completely missing.

The results for the statement "IQ2 - It felt like the other participants and I shared the same place while we interacted in the APP.", where VR was also perceived as the clear winner, can be explained in a similarly. By characterizing the virtual environment with 3D objects and depicting oneself with an avatar, the feeling of sharing the same space with others is conveyed more strongly. This in turn also strengthens the awareness of the perception of the virtual space and changes occurring in it (IQ3). Therefore, regarding "IQ3 - I was always aware of what the participants were presenting or editing during the session." VR was perceived most positively. However, BBB takes second place here, since notifications and visual feedback elements alert users of BBB to changes in the system status. Finally, statement IQ4 asked whether the interaction options are appealing to exchange ideas and artifacts. From this point of view, too, VR was felt to be more beneficial than BBB and "Spatial". This emphasizes the importance of 3D representation of artifacts for the exchange of ideas, allowing for better content delivery and visual representations.

With a view to the research question (RQ3), it can be summarized that VR offers many advantages over conventional collaboration tools in terms of interaction properties and thus enables realistic communication that is closer to usual face-to-face events. Here, too, the hypothesis can be confirmed.

\subsection{Learning Support}

\begin{figure}
\centering
    \includegraphics[width=0.9\linewidth]{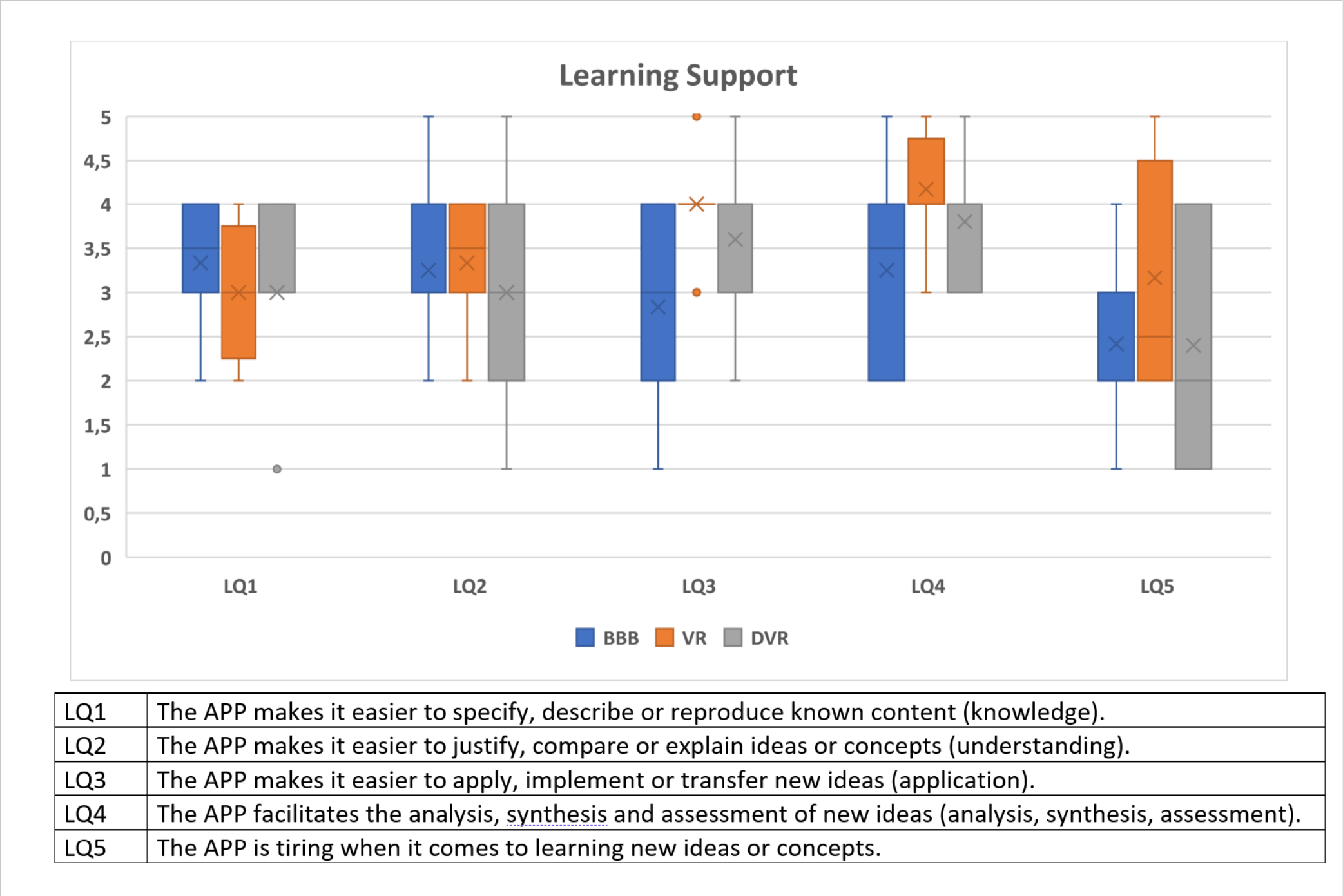}
    \caption{Evaluation results concerning interaction}
    \label{fig7}
\end{figure}

In order to analyze the perception of the learning support provided by the respective media (BBB, VR, and DVR), the participants in the seminar were asked for their assessment of the learning support of the taxonomy levels according to Bloom \cite{bloom1956handbook}. As shown in Figure \ref{fig7}, there were a total of five questions LQ1-LQ5. The LQ1 question aims to have the participants' perception of the learning conduciveness of taxonomy level 1 (knowledge) evaluated. Similarly, in LQ2 and LQ3, taxonomy levels 2 (understanding) and 3 (application) are asked about their learning support. In LQ4, the higher taxonomy levels 4-6, i.e. analysis, synthesis, and evaluation, were combined in one question and asked about their ability to promote learning. Finally, the final question LQ5 addresses how tiring the medium is to learn new concepts and ideas. LQ5 was asked intentionally negated in order to explicitly examine the cognitive load of the respective media.

Looking at the results in Figure \ref{fig7}, it becomes clear that specifying, describing or reproducing known content is perceived as easier on average via BBB than in VR or DVR. This is due to the fact that many well-known features such as chat function, screen sharing, freehand drawing and shared notes are possible in BBB, which simplify the presentation and transfer of known knowledge. In the "Spatial" collaboration platform used for VR/DVR, some of the features such as chat functions or screen sharing were not possible.

However, with regard to the support of the learning conduciveness of taxonomy level 2 (understanding), it becomes apparent that VR was perceived slightly more positively on average than BBB and DVR. This can be explained with the three-dimensional VR environment, which facilitates the understanding and comparison of different concepts through the combination of different multimodal media objects such as 3D objects, images, audio, video, etc. (see Figure \ref{fig3}).
VR was also rated significantly more positively than DVR and BBB with regard to the support of the learning conduciveness of taxonomy level 3 (application). This is strongly related to the interactivity of the VR learning environments, which make learning new ideas and concepts through movement and exploration naturally tangible and closer.

A similar trend can be seen with regard to the support of the learning conduciveness of the higher taxonomy levels 4-6 (analysis, synthesis and assessment). Again, participants felt that VR was, on average, more useful than BBB and DVR in analyzing, synthesizing and evaluating new ideas. On the one hand, this can be explained by the aforementioned point of interactivity and, on the other hand, by the additional factors such as fun and motivation, which have strengthened cooperation with VR and thus triggered a positive learning experience among the participants.

Looking at the results of the last question LQ5 regarding learning support, it can be seen that VR was rated as a tiring medium when it comes to learning new content or concepts. On average, VR also performs significantly worse than BBB and DVR. On the one hand, this can be explained by the nature of the VR headsets, which were not considered very ergonomic for many participants because they pressed on the nose and caused headaches during longer VR stays. On the other hand, the natural form of input control in VR via movement, gestures, etc. requires a higher expenditure of energy, which is not necessary when using BBB and DVR.

In summary, with a view to the research question (RQ4), it can be concluded that VR was not felt to be very conducive to learning support for the low taxonomy level 1. On the other hand, the learning conduciveness of the higher taxonomy levels, especially levels 4-6, was rated more positively than BBB and DVR due to the diverse possibilities for interactivity in VR.

\subsection{General Feedback and Group Work Logs}

In addition to the questionnaires described above, there were various open questions and group work protocols to collect general feedback from the participants on the VR seminar. These are summarized below and linked to the results described above.
In the group work protocols, the findings from the questionnaires regarding the perception of collaboration, interaction, and learning support were confirmed. It became clear several times in the minutes that the collaboration and the design of the VR presentation rooms were very time-consuming and the students overall had the impression that things were not running as efficiently and effectively as if they were creating or using a classic PowerPoint slide. Nevertheless, the notes of the group work protocols show that the participants had a lot of fun working together and that they were able to find free rein in their creativity in order to be able to present their ideas in a VR environment.
The open questions showed that, despite the very positive assessment of the interaction options in VR, it was difficult to make precise entries or type in longer text sequences on the virtual keypad. In addition, technical issues were mentioned where participants were thrown out of the VR room. This shows that a stable and fast internet connection must be ensured in order for the audio and graphic quality in VR to be adequate and to work properly despite the synchronization of larger 3D data models. In addition, some participants (2/6 of the VR participants) indicated that after longer stays in VR they experienced a feeling of discomfort related to the phenomenon of cyber sickness. In DVR, on the other hand, there were no complaints in this regard. Despite the suggestions for improvement mentioned, the majority opinion was that VR increases creativity and enables interactive exchange meetings despite physical distance.

\subsection{Student Feedback}

Every semester, students are asked about their opinion of the respective courses by a student event criticism (VKrit). For this purpose, the VKrit provides a written course evaluation from the faculty with a standardized evaluation form, which the students should fill out online. The VKrit is carried out anonymously for each course. Although the evaluation criteria of the VKrit are general and not tailored to specific events, some findings that are relevant to the research questions can be taken into account. The seminar did very well in the VKrit. Under the heading of overall impression, the participants in the survey voted "very good" (the highest grade on a scale consisting of five items from very good to very bad). Eleven of a total of twelve participants took part in the VKrit and provided their feedback. Many of the participants (especially the VR participants, who were provided with VR headsets) were positively impressed by the format and would like to see more events using VR technology: "It was very exciting to be almost exclusively in VR. Especially since the content of the seminar is closely linked to VR, the content and technology complemented each other very well. Even if face-to-face events are possible again, there should still be events like these that use VR as a medium." Some participants gave feedback that VR meetings are a nice alternative to video calls, but only for shorter meetings (up to 60 min) as they become strenuous from a certain point in time ("VR headsets press against the face"). Some of the participants wished that in the future all participants should be provided with VR headsets so that they can experience the event in full immersion.

\section{Findings}

This section summarizes various findings that were gained from the use of virtual reality in the computer science seminar. Different perspectives such as organization, collaboration, and interaction as well as learning success are discussed.

\subsection{Organization}

Due to the interactive discussion and workshop formats (idea generation, discussion, and presentation of ideas) at the organizational level and the direct content connection of the offered seminar topics to VR, a promising basis was given to assess the quality of VR as a medium for teaching and learning in the university context. However, from an organizational perspective, organizing the seminar using VR took significantly more time and effort compared to conducting a seminar in person or using traditional collaboration tools such as BBB or Zoom. This meant that participants had to be familiarized with the VR technology, including hardware and software. In addition, suitable virtual learning rooms had to be selected and designed for the respective sessions, which often involved a lot of preparation time. The reason for this is that the web interface of "Spatial" only offers limited possibilities for the precise design of the room and many details could often only be set and refined when entering the VR room via a VR HMD. Even if in some cases it is possible to reuse the learning rooms that have already been created in future learning scenarios, a cost-benefit calculation is very important in order to be able to precisely estimate the effort and thus be able to weigh up whether the use of the VR medium is really worthwhile brings additional profit. In addition to the design and implementation effort, which was also mentioned by the participants for the creation of the final presentation rooms, accessibility to the VR medium is an important aspect. While some of the participants were familiar with VR and had prior experience with Spatial, some participants were unfamiliar with some of Spatial's features. There was also the point that “Spatial” as a collaboration platform revealed some weaknesses that only became apparent when used with a larger group of participants. There were occasional breaks during the VR meetings, which occurred due to the connection quality of individual participants or because the VR space was getting larger and larger due to the objects created (text, graphics, 3D models, etc.) so scalability was not given during synchronization. Some of the features such as "spatial" audio or the general sound quality could still be improved and were often mentioned by the participants in discussions or in the group work reports. In the course of the seminar, there were also various updates to "Spatial", where changes were made to some interaction concepts, which the participants had to get used to again. It can therefore be observed that "Spatial" and many other VR collaboration tools are still in an experimental state and there are no technical standards on which such VR collaboration tools are based.

\subsection{Collaboration and Interaction}

Through the use of VR, the cooperation between the participants could be improved with regard to some factors. In particular, the results regarding the perception of presence show that social presence was able to greatly promote cooperation. On the one hand, this is related to the realistic representation of the participants through 3D avatars and, on the other hand, to the possibility of a realistic face-to-face interaction among the participants. This enabled a collaborative environment that was felt by many participants to be motivating for the learning process.

In addition to these positive insights, some negative insights made collaboration and thus joint learning more difficult. For example, collaboration in VR was rated poorer in terms of efficiency and effectiveness compared to conventional collaboration tools. This was also confirmed as feedback in the open questions and in the group work protocols.

\subsection{Learning Success}

The use of VR has shown that the teaching and learning experience can be enhanced through the increased perception of presence. The results regarding the perception of presence show, in particular, that VR has clear strengths here compared to conventional collaboration tools and can therefore positively promote the learning process, especially in group work. In addition, VR offers the possibility of multi-modal learning via different media artifacts (text, images, audio, video, 3D models, etc.), which appeal to different stimuli and senses. This makes it possible to address different types of learners equally. In addition, VR offers greater clarity and experience of learning content through virtual learning rooms, which is not supported in this way in conventional collaboration tools. These findings were also confirmed in the assessment of learning support, particularly with regard to the higher taxonomy levels. This illustrates the potential of VR to promote learning, especially in the application, analysis, synthesis, and assessment of knowledge.

\begin{figure}
    \includegraphics[width=1\linewidth]{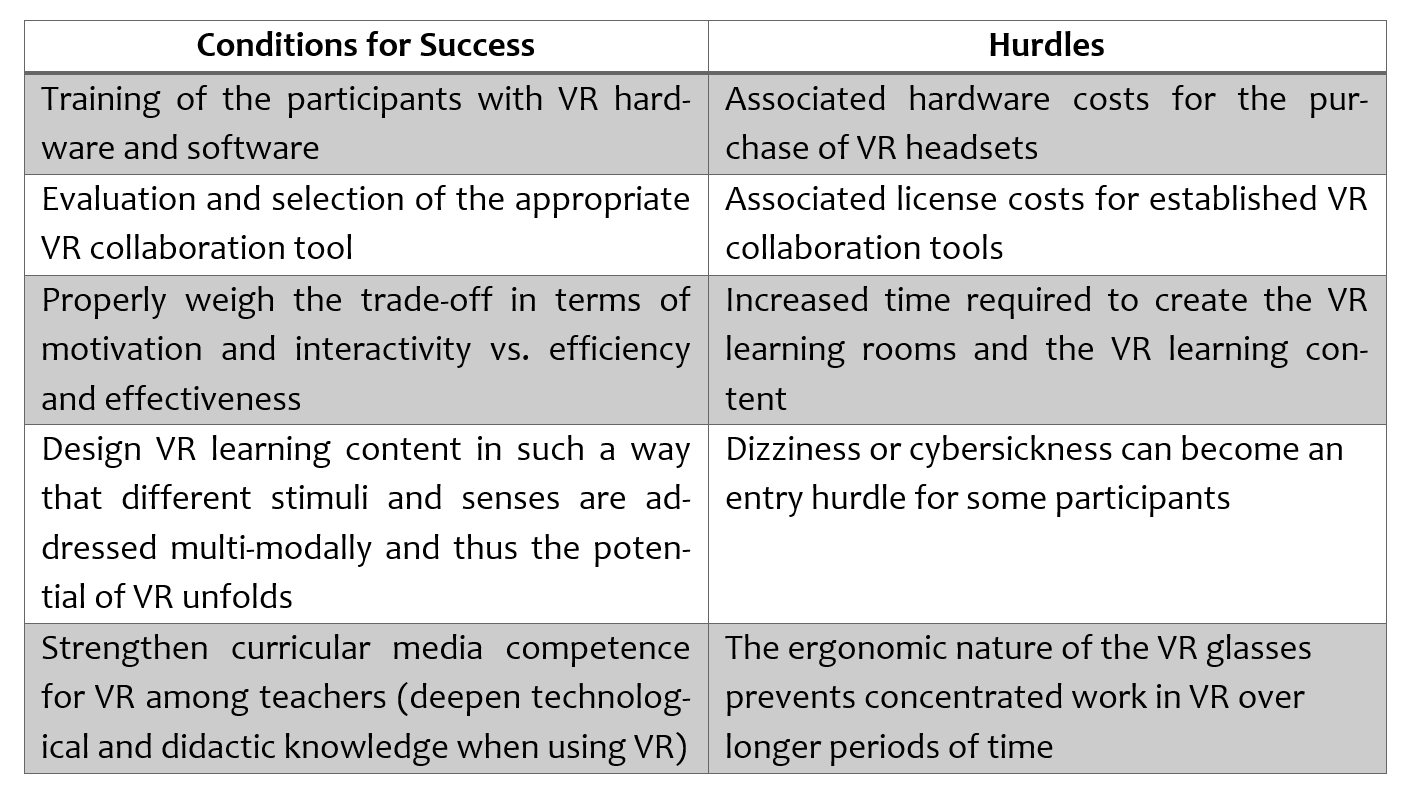}
    \caption{Conditions for success and hurdles for the use of VR}
    \label{fig8}
\end{figure}

In addition to the aforementioned opportunities, there are also some risks to learning success. The use of VR involves additional costs for purchasing the hardware. The results from Chapter 6 showed that an immersive experience in VR cannot be replaced by an on-screen experience on a non-immersive device class. In addition, some of the participants stated that VR caused dizziness and headaches, especially in longer sessions of up to two hours. This problem, also known as cybersickness (\cite{laviola2000discussion}, \cite{davis2014systematic}), occurs particularly in VR applications where a lot of fast and sudden movements happen. In addition, wearing a heavy HMD can also be seen as an obstacle to learning, since some of the participants also stated that the VR headsets can be ergonomically upgraded, that they press on the nose if worn for too long and cause a feeling of discomfort, which has a negative effect on concentration. In addition to these technical aspects, it was observed with regard to the learning success in the organization of the event that there is a general lack of teaching/learning concepts or a conceptual didactic basis for the use of VR. In addition, most teachers and learners do not yet have the curricular media competence for VR, so systematic training, introduction, planning, and use of the VR medium are required. Finally, the conditions for success and hurdles for the use of VR are briefly summarized in a tabular overview of Figure \ref{fig8}.

\section{Conclusion and Outlook}

Due to the constant digitization of learning and teaching media, educational institutions such as universities are faced with the challenge of finding a balance between the technical possibilities of the media and didactically meaningful use. Virtual reality is said to have great potential to increase the learning success of students and to positively enrich teaching. Since the use of virtual reality in university education and in particular its effective use in teaching has not yet been examined in depth and there are no concrete pilot projects in this context, a computer science seminar based on VR as the primary medium was carried out in this work. Plenary sessions were held for workshops, discussions, and presentations in VR. In addition, there was also group work that was carried out entirely in VR. Based on the course of the seminar carried out here in this article, it was possible to compare the use of VR with conventional collaboration tools such as BBB in terms of usability, presence, collaboration, interaction, and learning success.

The results show that VR hardware and VR collaboration tools are not yet very well known to many of the participants and that the VR collaboration tool used cannot yet keep up with conventional, mature collaboration tools such as BBB in terms of user-friendliness. Nevertheless, VR has the potential to increase the clarity and experience of learning content through the increased perception of presence. With regard to collaboration, it was found that VR still has a lot of potential for improvement in terms of efficient and effective collaboration, for example, to support the precise input and editing of virtual rooms. Nevertheless, the possibilities of interaction in VR and in particular the face-to-face interaction with other participants were perceived as very positive, which is the greatest advantage over conventional collaboration tools. In addition, the fun factor in connection with VR was also strongly emphasized, as many participants found the use of VR to be motivating for exchange and learning success.
It should not go unmentioned that there are limitations to this study. For example, the small number of VR participants should be mentioned. In addition, there is the aspect that some participants had previous experience in the field of VR, which could have had a positive effect on the results of the evaluation. It should also be mentioned that the course was based on "Spatial" and that there are other VR collaboration tools that could influence the analysis of the results. Seen from this perspective, the results presented here should be regarded as initial assessments. Overall, however, the knowledge gathered shows that the strengths of VR in terms of interaction and increasing motivation to learn can be used in a targeted manner to create benefits through the digital medium VR even in the post-corona period. VR is therefore not only to be seen as a medium that enables realistic interaction and communication for distance teaching, but also as an additional enrichment of the known teaching formats.

As an outlook, it can be summarized briefly that the developments in hardware and display technology are constantly advancing and give hope that the required hardware will become more immersive, ergonomic, and affordable so that an even broader use of this technology can be made possible. In addition, the latest software advances in the field of VR show that more and more context-sensitive learning aids are emerging that enable complex and dangerous learning scenarios and enable users to transfer learned (psycho)motor skills into reality. Due to the trend in the field of artificial intelligence, AI-supported methods are increasingly being used in VR-based learning and training environments in order to support the learning process even more effectively and efficiently. In this context, it will be even more important in the future to continuously analyze the effects of VR on teaching and learning processes and to use this medium in a targeted manner.

\bibliographystyle{splncs04}
\bibliography{references}

\end{document}